# Optical Pumping in Ferromagnet-Semiconductor Heterostructures: Magneto-optics and Spin Transport

A. F. Isakovic,[1] D. M. Carr,[2] J. Strand,[1] B. D. Schultz,[2]
C. J. Palmstrøm,[2] and P. A. Crowell[1]

[1]*School of Physics and Astronomy,* [2]*Department of Chemical Engineering and Materials Science, University of Minnesota, Minneapolis, MN 55455*

Abstract

Epitaxial ferromagnetic metal – semiconductor heterostructures are investigated using polarization-dependent electroabsorption measurements on GaAs p-type and n-type Schottky diodes with embedded $In_{1-x}Ga_xAs$ quantum wells. We have conducted studies as a function of photon energy, bias voltage, magnetic field, and excitation geometry. For optical pumping with circularly polarized light at energies above the band edge of GaAs, photocurrents with spin polarizations on the order of 1 % flow from the semiconductor to the ferromagnet under reverse bias. For optical pumping at normal incidence, this polarization may be enhanced significantly by resonant excitation at the quantum well ground-state. Measurements in a side-pumping geometry, in which the ferromagnet can be saturated in very low magnetic fields, show hysteresis that is also consistent with spin-dependent transport. Magneto-optical effects that influence these measurements are discussed.

PACS Numbers: 85.75.-d, 73.40.-c, 78.67.De

One of the outstanding problems in the physics of heterostructures is the transport of spin across an interface. There has been considerable progress in the case of all-semiconductor junctions, in which the spin in one material is polarized by an enhanced Zeeman interaction,[1, 2] the use of a p-doped ferromagnetic semiconductor,[3] or by optical pumping.[4] Until recently, there has been less dramatic success in the case of ferromagnetic metal-semiconductor junctions.[5] Although the existence of spin-polarized currents flowing from NiFe into a GaAlSb/InAs two-dimensional electron gas (2DEG) has been inferred from transport measurements,[6] there are a number of background effects that make the interpretation of these experiments difficult.[7-9] Furthermore, arguments have been proposed that place fundamental limitations on the efficiency of spin injection for diffusive transport at an ohmic metal-semiconductor contact.[10] Although most metal-semiconductor junctions are in fact not ohmic, the relative success of the all-semiconductor approach has reinforced the pessimistic view of the situation for ferromagnetic metals on semiconductors.

We report in this paper on the spin polarization of photocurrents induced by optical pumping of n-type and p-type Schottky diodes. These currents are due to minority carriers flowing from the semiconductor into a ferromagnetic metal. The samples were prepared by molecular-beam-epitaxy (MBE) growth of 200 Å films of Fe and $Fe_{0.5}Co_{0.5}$ on (100) $GaAs/In_{1-x}Ga_xAs$ heterostructures. Control samples with Al in place of the ferromagnet were also prepared. After growth of the semiconductor layers, each sample was cooled to 95 ºC before the deposition of the metal. Details of the growth procedure, which limits interfacial reactions and yields magnetic layers of high crystalline and magnetic quality, have been given elsewhere.[11] A description of the various samples is provided in Table I. Samples A, B, C, and E are Schottky diodes with $In_xGa_{1-x}As$ (x = 0.11) quantum wells embedded in the depletion region. All of the heterostructures were grown on either $n^+$ or $p^+$ ($10^{18}$ $cm^{-3}$) substrates. The quantum wells, comprising 100 Å of $In_{0.11}Ga_{0.89}As$ between undoped 1000 Å GaAs barriers, were grown on doped ($10^{17}$ $cm^{-3}$) buffer layers. Sample A includes an additional undoped 1000 Å GaAs spacer layer before the metal film. Samples A, B, C, and E show typical Schottky diode behavior with ideality factors less than 1.2. Samples D and F, which were grown on $p^+$ substrates, contain a 1200 Å n-doped layer ($10^{16}$ $cm^{-3}$) above the quantum well, followed by a Si ($10^{12}$ $cm^{-2}$) δ–doping layer, 50 Å of undoped GaAs, and the Fe layer. The transport behavior of the δ-doped samples is more complicated, since they comprise a p-n junction in series with a heavily doped Schottky contact. However, their photodiode characteristics when the metal-semiconductor contact is reverse-biased are similar to those observed in the simple Schottky structures.

The experimental technique is similar to that used by Prins *et al.*[12] and Hirohata *et al.*[13]. The samples were patterned into mesas using photolithography and wet etching. The experiments were generally performed on strips 200 μm wide and approximately 2 mm long. For the first set of measurements reported here, light from a helium-neon or Ti:sapphire laser was focused at normal incidence on the semitransparent metallic layer of the Schottky diode. The light was chopped at frequencies from 200 – 500 Hz, and its helicity was modulated between right and left at a frequency of 42 kHz using a photoelastic modulator. Photocurrents were detected at both the polarization modulation and chopping frequencies using two lock-in amplifiers. The ratio of these signals is the fraction ΔI/I of the total photocurrent that is sensitive to the polarization of the exciting light, and it is thus an upper bound on the spin-polarization of the photocurrent. We will refer to the ratio ΔI/I as the photocurrent polarization signal, and it is shown in Figure 1 as a function of magnetic field at two excitation energies above the GaAs band edge. These data, which are typical of all the structures we have studied, were obtained on sample A. Measurements were conducted as a function of magnetic field, excitation energy, and bias voltage. Forward bias voltages large enough to turn on the diodes produced phase shifts in the photocurrents and hence erroneous results. The temperature was fixed at 10 K for the measurements discussed here.

The data shown in Fig. 1 track the magnetization of the overlying $Fe_{1-x}Co_x$ film very closely, particularly at the highest excitation energy. This fact is demonstrated in the inset of Fig. 1, which shows the polarization signal superimposed on the polar Kerr effect data measured on sample A. The polar Kerr rotation angle $\theta_k$, which is proportional to the magnetization, has been scaled to match the polarized photocurrent signal at high field. An important question is whether the measured signal reflects a true spin-dependent current, which would be sensitive to the spin-polarized density of states in the ferromagnetic layer, or if it is due primarily to the magneto-absorption in the ferromagnetic film. To address this issue, we polished a piece of sample A to a thickness of approximately 50 microns and mounted it so that the magneto-absorption could be measured in transmission. The magnetoabsorption measured at E = 1.45 eV is shown in Fig. 1 as the solid curve. It is evident that the magnetoabsorption and the photocurrent polarization signal are comparable in magnitude, so that at most ~0.5 % of the photocurrent signal can be attributed to a true spin-dependent current.

Given the large background contribution from the ferromagnetic film, we have investigated other means of isolating a true spin-dependent transport signal. One approach is to vary the reverse bias voltage. Subtraction of the polarized photocurrent data obtained at different biases removes the background magneto-optical effect and thus provides a lower bound on the real transport polarization signal. We have carried out this procedure for energies from 1.95 eV down to the GaAs band edge. The difference of the polarization signal measured at 1.5 V and 0 V is approximately -0.5% and also tracks the magnetization of the ferromagnetic film.

A much larger photocurrent polarization signal, up to 35 % at 5 T, is seen in Fig. 2a, which shows spectra at different magnetic fields for excitation energies near the quantum well ground-state in sample A. These spectra were obtained at small forward bias, so that the bands near the quantum well were nearly flat. The photocurrent, however, still flows in the reverse direction due to the band-bending in the Schottky region. The electroabsorption spectrum, which is simply the raw photocurrent signal normalized by the incident intensity, is shown in Fig. 2b. The weak peaks fanning out at higher fields in the electroabsorption spectrum are attributed to Landau levels, each of which produces a corresponding derivative feature in the photocurrent polarization signal. The polarization signal has a weak background with essentially the same field dependence as observed at higher excitation energies, and we attribute this primarily to the magneto-absorption in the ferromagnetic film discussed above. Unlike the data obtained near and above the GaAs band-gap, both the polarization and raw electroabsorption spectra depend very strongly on the reverse bias voltage, which produces a substantial Stark shift. Both signals are smaller in the reverse bias regime.

We focus our discussion here on the ground-state peak at 1.42 eV in Fig. 2b and the corresponding feature in the photocurrent polarization signal, which approximates the derivative lineshape expected for a typical magnetic circular dichroism (MCD) spectrum. The magnitude of the largest polarization signal, obtained from the peak at 1.416 eV in the case of sample A, is shown as a function of magnetic field in Fig. 3a for all of the samples in this study. Unlike the field-dependence measured above gap, it is clear that these data do not simply track the magnetization of the overlying film. This is not surprising given that the narrow peak in the electroabsorption spectrum makes the absorption coefficient for a given polarization extraordinarily sensitive to the Zeeman splitting of the quantum well states. [14] In the case of a single Zeeman-split level, we expect the corresponding MCD to be linear in applied field, and we begin our analysis by fitting each of the data sets in Fig. 3a to a straight line above 3 T and subtracting the fit from the entire data set. The results are shown for each sample in Fig. 3b.

The data of Fig. 3b still contain a background contribution from the magneto-absorption in the ferromagnetic film, which is of order 2 % at 5 T for samples A – D. Furthermore, we cannot claim to have eliminated all possible contributions to the background from magneto-optical effects in the quantum well. Nonetheless, there are several important features of these data. The p-type samples show the largest signals, and the two smallest signals are observed from the n-type Sample C and the completely non-magnetic control sample E. In fact, the subtracted signal observed in Sample C is consistent with the magneto-optical background due to the Fe film, and control sample E shows zero signal within our error limits. The signals saturate at fields from 2 – 2.5 T, which is the same range of magnetic saturation fields observed in polar Kerr effect experiments. These observations are consistent with expectations for spin-dependent transport from the quantum well to the ferromagnet. In the case of p-type samples, the photocurrent is carried by electrons, for which observed spin lifetimes are orders of magnitude longer than for holes.[15]. If the time for the photo-excited carriers to escape from the quantum well is comparable to or longer than the spin lifetime, then the spin polarization of the photocurrent should be correspondingly reduced. For this reason, one expects a larger polarization signal in the p-type Schottky diodes.

Although we believe that some of the trends observed in Fig. 3b are evidence for spin ejection from the semiconductor into the ferromagnet, the behavior of control sample F is not consistent with this hypothesis. This sample was prepared from sample D by etching away the ferromagnetic layer and then evaporating Al in its place. The quantum well is thus exactly the same as for sample D, and, with the caveat that the metal was grown *ex situ*, this allows for an essentially ideal control. The corrected signal of Fig. 3b is significantly smaller for sample F than for sample D but is still noticeably greater than zero. An obvious question in this case is whether any Fe was left at the interface after etching, a problem which would not have affected the other control sample (E), in which the non-magnetic layer was prepared *in situ.* More generally, however, the difficulties in interpreting these measurements revolve around the large fields required to saturate the ferromagnetic film in the perpendicular direction. These are responsible for the large magneto-optical effects observed in the semiconductor.

The side-pumping geometry shown in the inset of Fig. 4 addresses this consideration. Since the in-plane coercivities are on the order of 20 Oe along the easy direction,[11] the magnetization can be switched in fields that are negligible on semiconductor scales. Furthermore, the light enters the cleaved facet of the sample without passing through the ferromagnetic film. As can be seen in Fig. 4, the polarized photocurrent signal, measured here on sample D, clearly traces out the hysteresis loop of the ferromagnetic film when the system is excited below the GaAs band edge. Although the signal has a broad maximum near 1.4 eV, we do not find any enhancement of the polarization signal at the quantum well ground-state exciton, in contrast to the case of normal incidence. This observation is a consequence of the fact that the heavy-hole is strongly confined, and so its spin is locked along the growth direction by the QW potential.

Although the existence of a magneto-optical effect, possibly due to successive reflections from the GaAs-metal interface as light propagates in the interior of structure, cannot be excluded, the data of Fig. 4 are suggestive of spin-dependent transport. As with the other data presented in this paper, a more conclusive statement would require a detailed microscopic understanding of the photocarrier transport at the metal-semiconductor interface. Since there is no barrier for the minority photocarriers, the physical situation is similar to majority-carrier transport in a forward-biased diode, which does not depend on the density of states in the metal if the transport is completely diffusive.[16] Hence, the observation of spin-dependent transport requires that the transport is at least in part ballistic, although the measurements introduced here cannot address this quantitatively.

In conclusion, a systematic study of polarization-dependent photocurrents in several ferromagnet/semiconductor Schottky diode structures has provided three important observations on the relationship between the optical pumping process and spin transport.

First, for excitation at normal incidence and photon energies above the GaAs band edge, there is a strong correlation between the observed polarization signal and the magnetization of the ferromagnetic film, but true spin transport is masked by large magneto-optical effects. These can be eliminated in part by measuring the bias-dependence of the photocurrent polarization signal. Second, much larger polarization effects are observed in spectroscopic and field-dependent measurements on $In_{1-x}Ga_xAs$ quantum wells placed in the depletion region. Although background contributions are also very large in this regime, p-type Schottky diodes show a polarization signal that tracks the field-dependent magnetization of the ferromagnetic electrode and is as large as 10% for excitation at the quantum well ground state. Finally, we have used a side-pumping geometry that allows for the detection of spin transport effects in very small in-plane fields, thus reducing background effects. A detailed understanding of these phenomena will be essential in identifying spin transport in ferromagnetic metal-semiconductor structures.

We acknowledge helpful discussions with Michael Flatté. This work was supported by DARPA/ONR-N/N00014-99-1005, ONR N/N00014-99-1-0233, NSF MRSEC 98-09364, and the Alfred P. Sloan Foundation (PAC).

**Figure Captions**

Figure 1: The polarized photocurrent signal ΔI/I as a function of the applied magnetic field at excitation energies of 1.50 and 1.95 eV is shown for sample A at T = 10 K. The differential magneto-absorption measured on the same sample is shown by the solid curve. The inset shows the polarized photocurrent data obtained at 1.95 eV (triangles) superimposed on the scaled polar Kerr rotation data. The scale factor is chosen so that the two data sets overlap at 4 T.

Figure 2: (a) The spectral dependence of the polarized photocurrent signal for sample A at several magnetic fields. The curves are displaced for clarity. (b) The corresponding electro-absorption data showing the large quantum well ground-state peak and a series of field-dependent peaks attributed to Landau levels, indicated by dashed lines.

Figure 3: (a) The maximum value of the polarized photocurrent signal at the ground-state exciton is shown as a function of magnetic field for all of the samples in this study. (b) The same data are shown after the subtraction of a background determined by fitting the data of (a) to a straight line above 3 T.

Figure 4: The polarized photocurrent signal is shown as a function of the magnetic field for sample D at T = 10 K in the side-pumping geometry illustrated in the inset. Excitation energies are indicated in each panel.

Table I: Details of the samples discussed in this paper. $\phi_b$ is the Schottky barrier height and $n$ is the ideality factor. QW stands for a quantum well of 1000Å (i)GaAs/100Å $In_{0.11}Ga_{0.89}As$/1000Å (i)GaAs, except for samples D and F, where it stands for 500Å (i)GaAs/100Å $In_{0.11}Ga_{0.89}As$/500Å (i)GaAs. Metal layer thicknesses are 200 Å.

| Sample | Heterostructure | Metal | $\phi_b$ (V), n |
|---|---|---|---|
| A | i(1000 Å)/QW/p/p+ | FeCo | 0.62, 1.12 |
| B | QW/p/p+ | Fe | 0.50, 1.15 |
| C | QW/n/ n+ | Fe | 0.80, 1.05 |
| D | i(50 Å)/ n δ-doped/QW/p/p+ | Fe | N.A. |
| E | QW/p/p+ | Al | 0.50, 1.15 |
| F | i(50 Å)/ n δ-doped/QW/p/p+ | Al | N.A. |

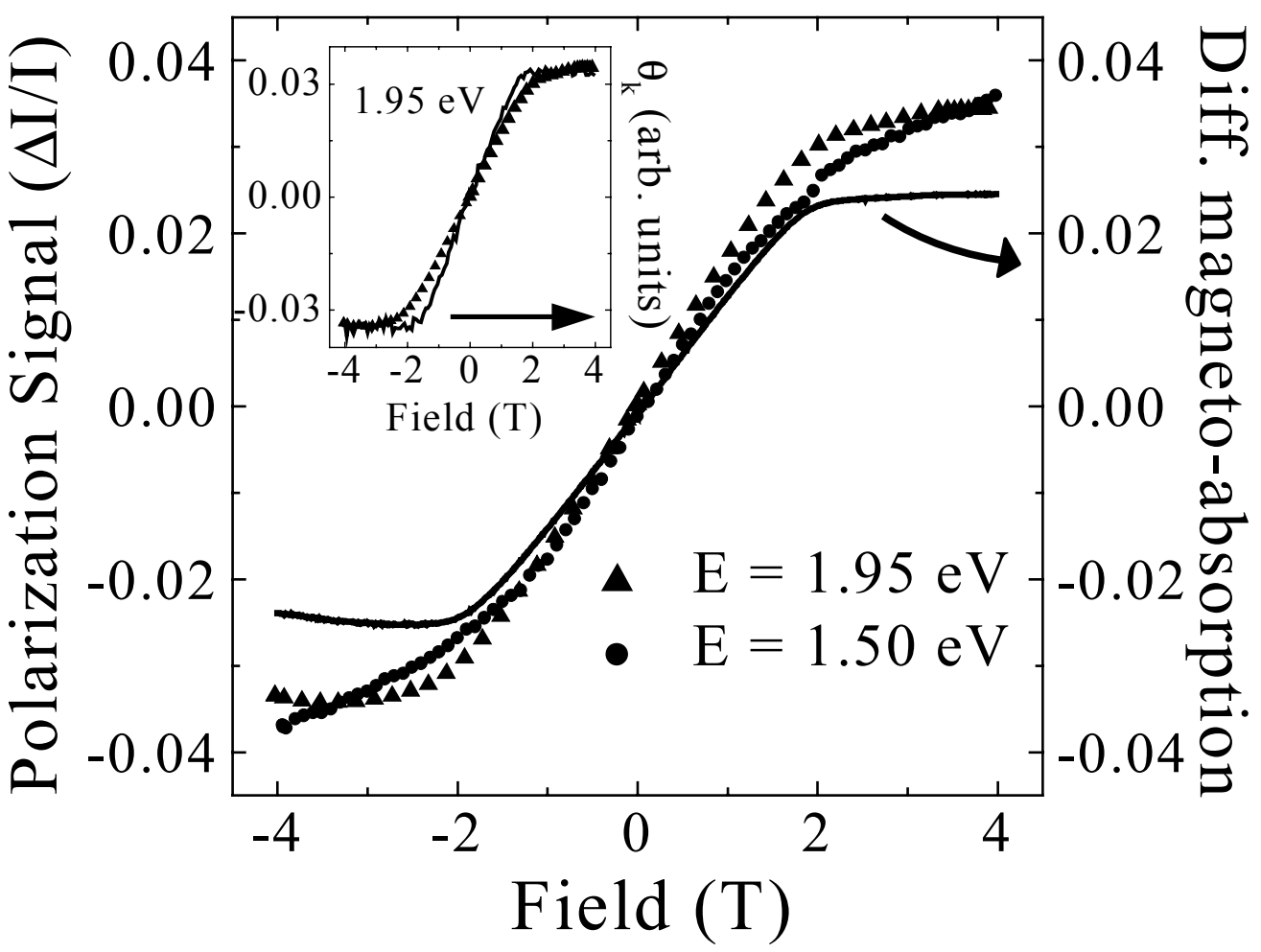

Fig. 1, Isakovic *et al*.

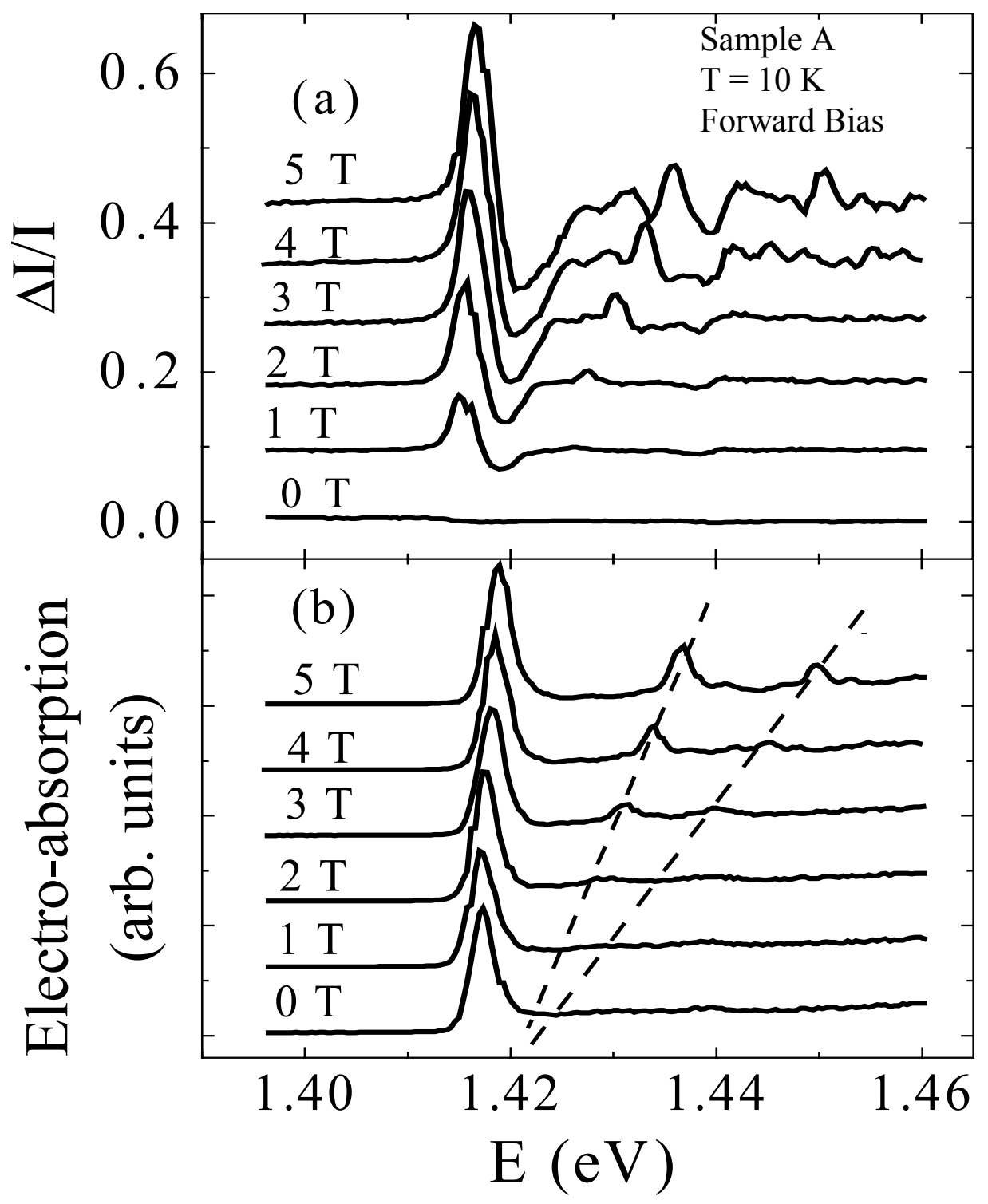

Fig. 2, Isakovic *et al*.

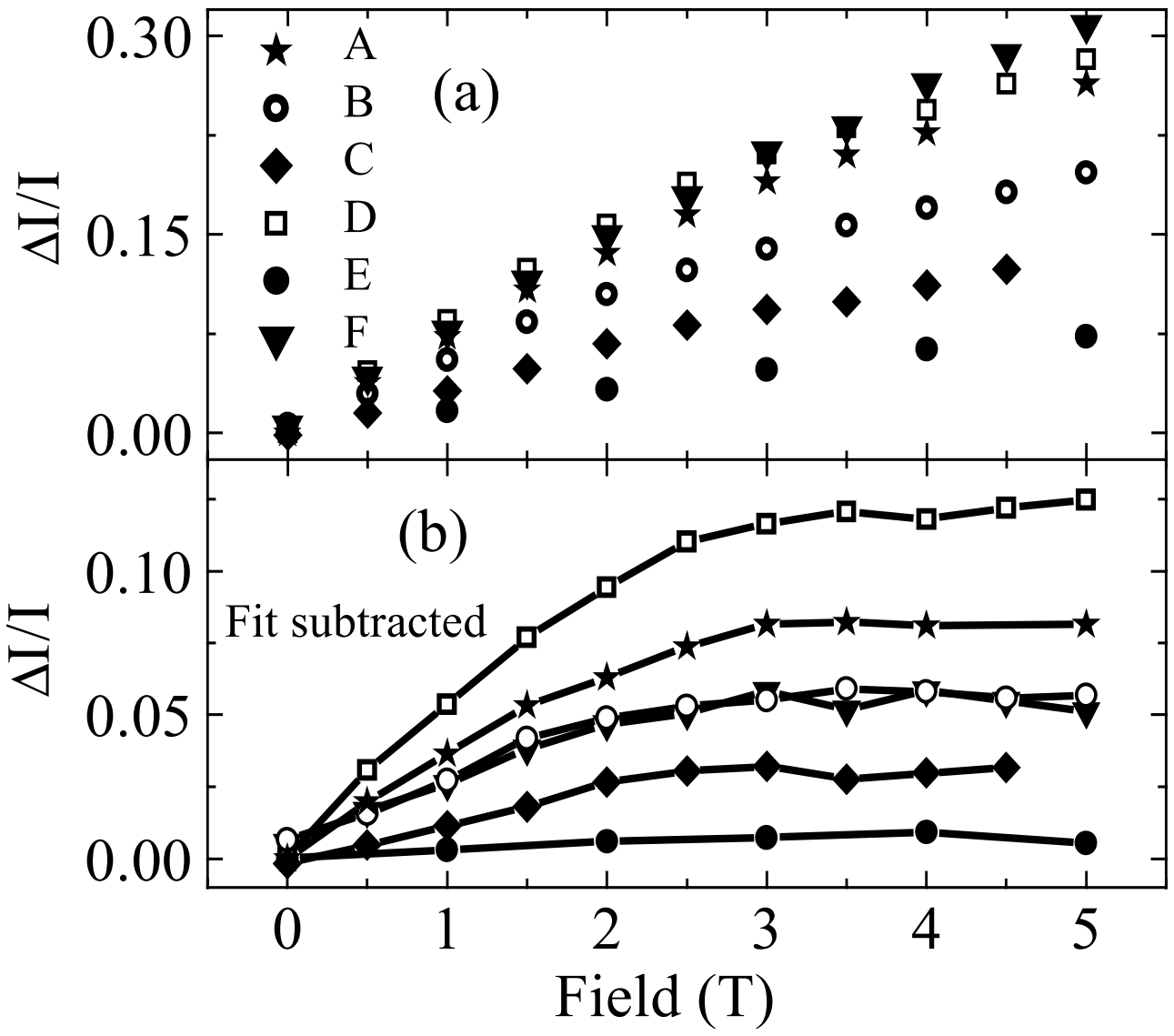

Fig. 3, Isakovic *et al*.

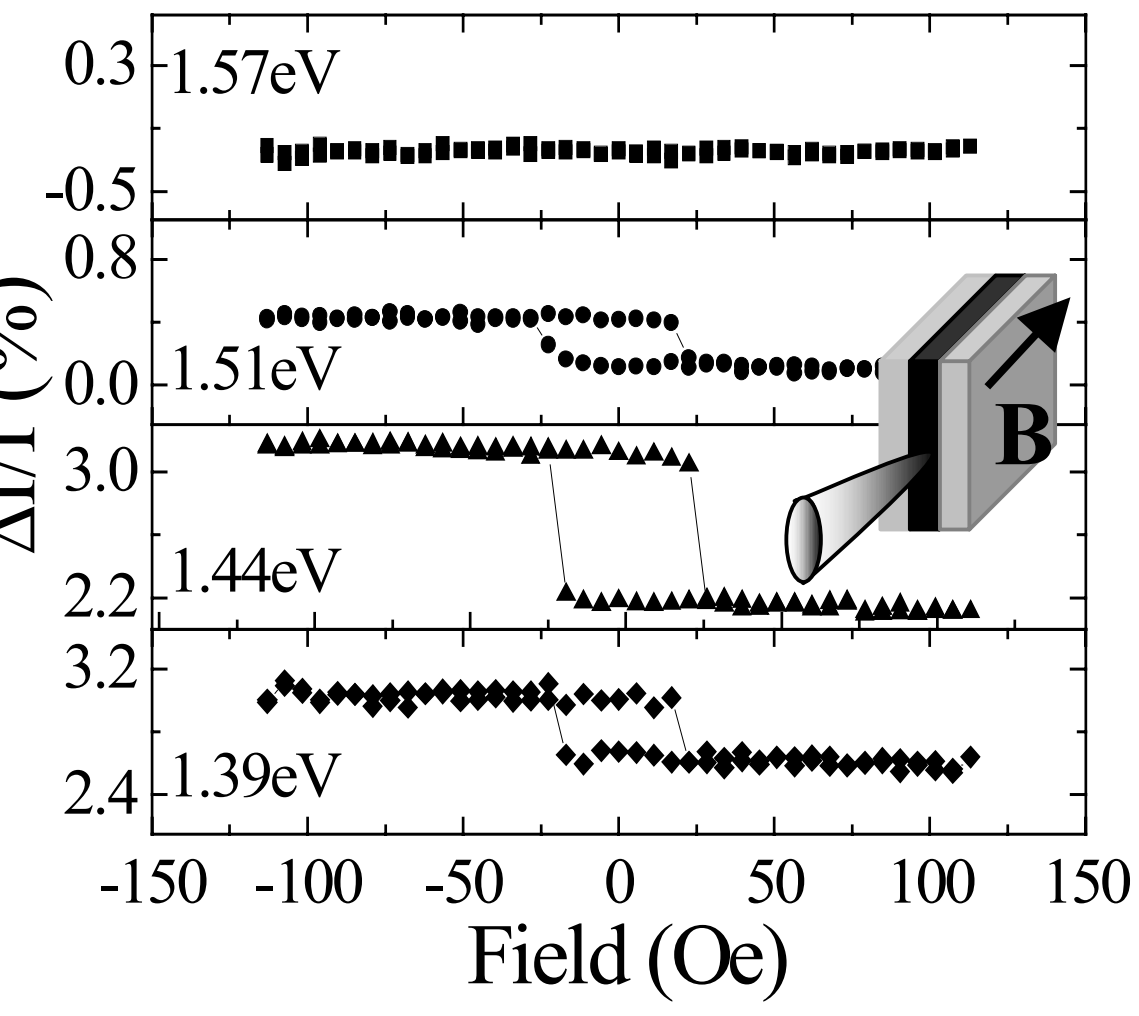

Fig. 4, Isakovic *et al*.